\documentclass{cup-pan}
\usepackage[utf8]{inputenc}
\usepackage{blindtext}

\usepackage[utf8]{inputenc}
\usepackage[T1]{fontenc}
\usepackage{graphicx}
\usepackage{wrapfig}
\usepackage{booktabs}
\usepackage{tablefootnote}
\usepackage{hyperref}
\usepackage{xcolor}

\usepackage{cite}

\usepackage{xurl}
\hypersetup{
    colorlinks,
    linkcolor={red!50!black},
    citecolor={blue!50!black},
    urlcolor={blue}
}

\title{ChatGPT: Applications, Opportunities, and Threats}

\author[1]{Aram Bahrini*}
\author[2]{Mohammadsadra Khamoshifar }
\author[3]{Hossein Abbasimehr}
\author[1]{Robert J. Riggs}
\author[4]{Maryam Esmaeili}
\author[5]{Rastin Mastali Majdabadkohne}
\author[6]{Morteza Pasehvar}

\affil[1]{Department of Systems and Information Engineering, University of Virginia, Charlottesville, VA, USA \newline *\url{ab4pn@virginia.edu}} 
\affil[2]{School of Computer Engineering, Iran University of Science and Technology, Tehran, Iran}

\affil[3]{Faculty of Information Technology and Computer Engineering, Azarbaijan Shahid Madani University, Tabriz, Iran}

\affil[1]{Department of Systems and Information Engineering, University of Virginia, Charlottesville, VA, USA}

\affil[4]{Department of Industrial Engineering, Faculty of Engineering, Alzahra University, Tehran, Iran}

\affil[5]{Department of Medicine, Faculty of Medical Sciences, The American University, Managua, Nicaragua}

\affil[6]{Department of Industrial Engineering, Faculty of Engineering, Islamic Azad University, Parand Branch, Tehran, Iran}

\corrauthor{Aram Bahrini}

\usepackage{tikz,eso-pic}

\AddToShipoutPicture{%
\begin{tikzpicture}[remember picture, overlay]
      \node[minimum width=4in] at ([yshift=-1cm]current page.north)  {Preprint accepted in IEEE Systems and Information Engineering Design Symposium (SIEDS) 2023};
    \end{tikzpicture}%

}

\begin{document}
\maketitle
\begin{abstract}
Developed by OpenAI, ChatGPT (Conditional Generative Pre-trained Transformer) is an artificial intelligence technology that is fine-tuned using supervised machine learning and reinforcement learning techniques, allowing a computer to generate natural language conversation fully autonomously. ChatGPT is built on the transformer architecture and trained on millions of conversations from various sources. The system combines the power of pre-trained deep learning models with a programmability layer to provide a strong base for generating natural language conversations. In this study, after reviewing the existing literature, we examine the applications, opportunities, and threats of ChatGPT in 10 main domains, providing detailed examples for the business and industry as well as education. We also conducted an experimental study, checking the effectiveness and comparing the performances of GPT-3.5 and GPT-4, and found that the latter performs significantly better. Despite its exceptional ability to generate natural-sounding responses, the authors believe that ChatGPT does not possess the same level of understanding, empathy, and creativity as a human and cannot fully replace them in most situations.\\
\end{abstract}
\section{Introduction}

Did you know that you could be talking to a machine without even knowing it? You might just be doing that with ChatGPT, an advanced natural language processing model developed by OpenAI, a research company co-founded by Elon Musk, Sam Altman, Greg Brockman, Ilya Sutskever, and John Schulman \cite{111}. ChatGPT is an advanced natural language processing (NLP) model that is trained on a vast amount of data \cite{Kirmani}, including billions of web pages and documents, making it capable of generating human-like text responses to prompts \cite{BBB}. Since its launch, it has quickly become one of the fastest-growing consumer applications in history, with an estimated 100 million active users monthly. Its vast knowledge base and language processing capabilities have the potential to revolutionize the way people interact with technology, making it easier and more natural to communicate with machines. However, while ChatGPT has impressive language processing capabilities and is an exciting technology with a wide range of potential applications in various fields, it still has limitations and challenges, including bias and the occasional generation of nonsensical output, known as ``hallucination'' \cite{DDDD}.

In this paper, we review the existing literature, investigate the main domains of ChatGPT, discuss the opportunities and threats it has on society, and explore its academic influence. We provide insights that are extendable to other systems facing similar challenges. 

\section{Background and Related Works}
The architecture known as GPT, initially introduced by OpenAI in 2018, serves as the basis for ChatGPT. The first version, GPT-1, had 117 million parameters to work with and was trained on a vast amount of text data obtained from the internet by utilizing a deep learning technique known as transformers. GPT-2, released in February 2019, improved substantially and had 1.5 billion parameters. OpenAI decided not to make the full version of GPT-2 available (only 8\% of the original model's size) to the public because of worries surrounding the model's potential for inappropriate use \cite{Hao}. GPT-3 was released with 175 billion parameters in June 2020, with a waitlist removed later in November 2021. It had advanced to version 3.5 by the time ChatGPT went public in November 2022. In March 2023, OpenAI made GPT-4 available to users who signed up for the waitlist and ChatGPT Plus subscribers in a limited text-only capacity. Nonetheless, it can respond to both text and images. Despite limited availability, GPT-4 has garnered attention for its improved performance compared to its predecessor. The reason for its superior performance over GPT-3.5 is that it has a larger model with more parameters tweaked during training in a neural network. To date (as of April 2023), OpenAI has not provided any information regarding the data, computing resources, or training techniques used to develop the language model \cite{FFFF}. OpenAI plans to release GPT-5 in November 2023.

Both theoretical and empirical studies have contributed to the development of ChatGPT. While empirical studies may be more prevalent in the literature, theoretical studies have also played a crucial role in advancing the field of natural language processing and deep learning.
The theoretical aspect of ChatGPT development involves the development of the underlying mathematical and computational models that enable the model to learn and generate human-like language. Numerous large language models have been developed in recent years, such as Bidirectional Encoder Representations from Transformers (BERT), XLNet, ChatGPT, and BLOOM \cite{Radford}. With the help of a single pre-training and fine-tuning pipeline, all of these transformer-based models can complete a variety of natural language processing tasks \cite{Kasneci}. These advancements offer significant potential in research and industrial contexts, and future developments are expected to lead to even more improved capabilities \cite{Scao}.

On the other hand, empirical studies of ChatGPT involve testing the model's performance in various NLP tasks, such as text generation, question answering, and language translation. A recent (March 2023) technical report published by Open AI on GPT-4 showed that the post-learning alignment process, a pre-trained Transformer-based language model, improved factuality and alignment with desired behavior and enables GPT-4 to perform at a human level on various professional and academic benchmarks, such as a simulated bar exam \cite{OpenAI}. There are many publications that study different aspects of ChatGPT, and we will review some of them in the next section summarizing the main applications, opportunities, and threats of ChatGPT.
This study sheds light on how ChatGPT can be used in different domains for the greater good while minimizing potential harm and how society can prepare for the opportunities and challenges presented by this emerging technology.

\section{ChatGPT Applications, Opportunities, and Threats}
Recent studies have explored various applications of ChatGPT in scientific literature. In this section, we choose the most important applications that can be used by ChatGPT and classify them into 10 different domains. These domains represent the broad range of possible topics that ChatGPT can assist with, and they are not exhaustive. We provide the applications, opportunities, and threats for each domain and study and provide detailed explanations of the first two domains because of their importance. 

\subsection{Business and Industry}
The business and industry domain encompasses a wide range of applications in operations and management, supply chain management, business analytics, transportation, human resources, marketing, e-commerce, accounting, finance, retail, real estate, and insurance. The main opportunities are increased efficiency and cost savings, improved decision-making and risk mitigation, better prediction accuracy and optimized planning, reduced workload, and fraud detection. For example, for supply chain management, ChatGPT can help with the following tasks:
\begin{itemize}
\item \textit{Demand forecasting:} ChatGPT can analyze historical data and market trends to predict demand for a product or service, which can help organizations optimize their inventory management and production planning decisions.

\item \textit{Inventory optimization:} ChatGPT can help businesses analyze their inventory data and suggest ways to optimize inventory levels, reduce stock-outs, increase efficiency, improve order fulfillment times, and save costs.

\item \textit{Supplier selection and evaluation:} ChatGPT can assist in identifying and evaluating potential suppliers based on factors such as quality, price, lead time, and reliability.

\item \textit{Logistics optimization:} ChatGPT can help with route optimization, carrier selection, and other logistics tasks to ensure the timely and efficient delivery of goods.

\item \textit{Risk management:} ChatGPT can analyze supply chain data to identify potential risks and provide recommendations to mitigate them, such as identifying alternate suppliers in case of disruptions to the supply chain.
\end{itemize}

The listed opportunities are just a few examples of how ChatGPT can assist with supply chain management tasks. It has the ability to process and analyze large volumes of data quickly, which can help businesses and industries make better decisions and improve their operations.
By leveraging these applications, they can achieve a range of benefits and stay competitive and thrive in today's fast-paced business environment.
It is essential to be mindful of the potential risks associated with any opportunity. ChatGPT can be vulnerable to certain threats impacting its effectiveness and usefulness. The main threats are dependence on available and high-quality data, unreliable and biased results,  hallucinations, lack of transparency, ethical and data privacy concerns, and cybersecurity risks. For example, in supply chain management, ChatGPT is unlikely to be a direct threat, but there are some potential concerns that businesses should be aware of when using this AI, which are listed as follows:

\begin{itemize}
\item \textit{Overreliance on technology:} There is a risk that the supply chain becomes overly reliant on ChatGPT's forecastings and recommendations, leading to a loss of human judgment and decision-making. The supply chain could be adversely affected if ChatGPT's results are inaccurate or biased.

\item \textit{Cybersecurity risks:} For ChatGPT to work, large amounts of data need to be available, which can be vulnerable to cyber attacks. Hackers could use ChatGPT to steal information and manipulate supply chain data if they gain access to it.

\item \textit{Data bias:} Accurate predictions and recommendations from ChatGPT are crucial for the success of the supply chain. To achieve this, all parties involved must ensure that the data used to train ChatGPT is unbiased and complete. To mitigate the risks associated with biased or incomplete data, it is essential to monitor data sources for potential biases regularly, establish clear protocols for identifying and addressing inaccuracies, and ensure that all stakeholders have access to accurate and comprehensive data sets.

\item \textit{Lack of transparency:} The complex algorithms behind ChatGPT make it difficult to understand how it makes predictions or recommendations. Consequently, a supply chain may not be able to evaluate ChatGPT's performance or verify its results due to a lack of transparency.

\item \textit{Ethical concerns:}  Concerns about how AI may affect jobs, privacy, and other ethical issues are present as ChatGPT is more integrated into supply chain management. Businesses must carefully consider the potential adverse effects of ChatGPT on their stakeholders, and steps should be taken to mitigate them.
\end{itemize}

Therefore, to mitigate the mentioned threats, businesses should maintain a healthy balance between human judgment and technology, put strong cybersecurity measures in place to protect their data against potential attacks, have appropriate data governance policies, use experts to analyze the outcomes, and carefully assess ChatGPT's fairness and accuracy.

\subsection{Education}\label{subedu} 
The education domain offers a wide range of technology tools and software applications that can be used for online learning, language learning, academic and general research, teaching assistance, writing, exam evaluations, feedback on assignments, educational content creation, professional development, and improving students' educational experiences. The main opportunities are increased access to education, material support for teaching and personalized study materials, fostering greater inclusivity (by providing language support, personalized learning experiences, diverse and inclusive content, and accessibility features), improved learner engagement, support for assessment such as efficient grading and customized feedback, and improvement with learning, such as in science, programming, languages, writing abstracts, essays, grant proposals, and application letters. ChatGPT can be a valuable tool in fostering diversity, equity, and inclusion in education by offering diverse perspectives and resources, promoting equitable learning, and facilitating dialogue and cultural competency \cite{Michels, 12, Dennison}. For example, in the applications of academic and general research and teaching assistance, ChatGPT can help with the following tasks:

\begin{itemize}
\item \textit{Teaching Assistance:} ChatGPT can produce instructional content and create presentations by providing templates, generate diverse perspectives and facilitate discussions on diversity and inclusion, suggesting images and graphics, and recommend ways to make the presentations more engaging. It can create quizzes or tests to assist with teaching, help evaluate assignments, projects, and exams, and provide feedback to students and teachers, although it cannot grade actual exams. It can also answer common questions, help understand complex concepts, and provide sample examples for programming languages such as R, Python, Java, and many more.

\item \textit{Research Assistance:} ChatGPT can assist with suggesting research ideas and methodologies (whether qualitative or quantitative) and providing examples of how these methodologies have been used in previous studies. It can also help enhance inclusivity in research, find the relationship between two subjects of interest, do statistical analysis and data interpretation, summarize the main contribution, and suggest further study extensions. These can be obtained by giving the article's information or copying and pasting the study into the chatbot.

\item \textit{Writing Assistance:} ChatGPT can help improve writing skills by providing feedback on sentence structure, grammar, vocabulary, punctuation, citations, and plagiarism. It can also suggest ways to organize ideas and brainstorm, make arguments more compelling, provide examples of writing styles, and make sentences shorter or longer.
\end{itemize}

The use of ChatGPT in education should be responsible and ethical, and ongoing research and observation are essential \cite{Hao}, as it could directly affect the learning experience. Implementing AI in education can lead to various ethical and societal threats. These include perpetuating systemic bias and discrimination, infringement of students’ privacy, escalation of student monitoring and surveillance, undermining student autonomy, and marginalizing traditionally underrepresented students. Furthermore, introducing AI in education can amplify various forms of inequity, such as racism, sexism, and xenophobia \cite{Akgun, Trust}.
The main threats of ChatGPT in education include over-reliance on technology and dependence, inaccurate or biased information, ethical concerns such as plagiarism, privacy, and misuse (generating fake news or spreading misinformation), lack of human interaction and decreased motivation, inability to handle certain tasks because of lack of understanding of the context, technical failures such as glitches and server downtime, and security risks. For example, in the applications of academic and general research and teaching assistance, the threats are listed as follows:
\begin{itemize}
    \item \textit{Decreased creativity and critical thinking due to over-reliance on technology:} Over-reliance on ChatGPT by researchers, teachers, and students may lead to a decrease in creativity, critical thinking skills, as well as a lack of diversification in research methods. Students may become less motivated to learn if they feel that ChatGPT is doing the work for them.
    
    \item \textit{Inaccurate or biased information:} ChatGPT's responses may not always be accurate, and it may unintentionally perpetuate biases and reinforce stereotypes in its training data. This could lead to erroneous conclusions or perpetuate existing biases in the research and misinform students and lead to incorrect conclusions.

    \item \textit{Lack of human interaction and decreased motivation in teaching:} While ChatGPT can provide assistance and feedback, it cannot replace the value of human interaction and personalized feedback from a teacher or mentor, which is essential for students' social and emotional development.

    \item \textit{Lack of transparency and difficulty in handling complex tasks in research:} It can be sometimes difficult to understand how ChatGPT arrives at its responses, which can make it hard to validate its findings or identify errors or biases. While ChatGPT can handle a wide range of tasks, it may not be able to handle more complex tasks or understand the context of certain tasks, which may limit its usefulness in some research contexts.

    \item \textit{Technical issues:} Like any technology, ChatGPT may face technical issues such as glitches, server downtime, or compatibility issues with certain software or data formats, which may disrupt the teaching and research process.

    \item \textit{Ethical concerns:} There may be ethical concerns related to using ChatGPT in academic research, such as privacy concerns or issues related to data ownership and control and consent. Also, students or researchers may misuse ChatGPT by copying and pasting information without proper citation, leading to plagiarism.

    \item \textit{Security concerns:} Storing sensitive data, such as student grades, test scores, and personal information, on ChatGPT could pose a security risk, thereby increasing the potential for data breaches.
\end{itemize}

To mitigate these threats, educators can use a variety of strategies to address the potential negative effects of ChatGPT, such as emphasizing critical thinking and analysis to counteract any decreased creativity and motivation, highlighting the importance of critical thinking and analysis, supplementing ChatGPT with other education methods, ensuring that students are aware of the potential for inaccurate or biased information, promoting collaboration and peer learning to address the lack of human interaction, establishing clear guidelines and policies for using AI technology to address ethical and security concerns, and having backup plans in place to address any technical issues that may arise.

\subsection{Science and Technology}
The science and technology domain has applications in modeling, artificial intelligence (such as machine learning, deep learning, and NLP), information technology (computing, software development, and data analytics), programming and coding, the Internet of Things (IoT), cryptography, cybersecurity, manufacturing, automotive, robotics, aviation, and energy.
ChatGPT presents numerous opportunities in various areas within the science and technology domain. Its applications in artificial intelligence include language translation, image recognition, and predictive modeling. It can also help facilitate secure and decentralized transactions in cryptocurrency. In cybersecurity, ChatGPT is useful for detecting phishing attempts, identifying potential security threats, and analyzing malware behavior. ChatGPT's ability to process large amounts of data is also beneficial in manufacturing, where it can help improve product quality, reduce costs, and enable better equipment maintenance. Complex automotive, robotics, aviation, and energy systems can also benefit from ChatGPT's data-driven analysis and decision-making capabilities. Overall, ChatGPT's advanced language processing and analytical capabilities make it a valuable tool in a wide range of science and technology applications.
The main threats are ethical, privacy, and security concerns; potential for biased recommendations; and potential for errors or misunderstandings.

\subsection{Government and Politics}
The government and politics domain has applications in politics and management, international relations, public administration, public safety, taxation, law and justice, contracts and agreements, and military.
The opportunities of ChatGPT in this domain can be classified as follows: in politics, it can assist with improved voter outreach, more efficient campaigning, and enhanced policy analysis; in public safety, it can help improve response times and optimize resource allocation; in taxation, it can facilitate individual tax filing; in law, it can assist with drafting lawsuits, improve the efficiency of legal research, enhance contract analysis, and streamline document review; and in the military, it can assist with improved intelligence analysis, logistics optimization, and more efficient training.
There are various threats related to the use of ChatGPT in this domain, including privacy and security issues, ethical concerns, the potential for errors or misunderstandings, and the possibility of biased recommendations.

\subsection{Healthcare and Medicine}
The healthcare and medicine domain has applications in healthcare systems, public health, mental health, biology, biotechnology, telemedicine, medical examination, pharmaceuticals, and veterinary medicine.
ChatGPT has many opportunities in this domain, and there are many interesting studies for applications in medicine, public health, and healthcare systems, which we refer the readers to check the studies in \cite{King, Kung, 16, 19, 21, 22, 23, 26, 27}. In healthcare, it helps with efficient triage and symptom checking, personalized health recommendations, and improved patient outcomes; in biotech, pharmaceutical, and veterinary medicine, it can improve efficiency, reduce costs, and improve patient outcomes or animal health. 
Regarding the threats in this domain, it presents a range of legal, regulatory, ethical, and privacy concerns. These concerns include data privacy and security issues, as well as the possibility of biased recommendations and errors in the system's output. 
\subsection{Infrastructure}
The infrastructure domain encompasses applications in construction, energy and water infrastructure, architecture, urban planning, and interior design. The main opportunities in this domain include improved efficiency, reduced costs, enhanced safety, and increased sustainability. However, there are also potential threats related to privacy and security concerns, the possibility of biased recommendations, and ethical concerns.
\subsection{Environment and Sustainability}
The environment and sustainability domain encompasses a range of applications, including waste management, renewable energy, climate change, agriculture, environmental policy, biodiversity, social equity, ecotourism, and food sustainability. This domain offers opportunities to improve efficiency, energy management, sustainability, reduce costs, and increase the utilization of renewable energy. However, it is crucial to be aware of potential threats such as cybersecurity risks to connected devices or systems, data breaches compromising privacy, misuse of data leading to biased analysis, and unintended environmental consequences of sustainability initiatives or technologies. By proactively addressing these potential threats, stakeholders in the environment and sustainability domain can work towards realizing the benefits of these applications while minimizing risks.
\subsection{Communication}
The communication domain has applications in media and entertainment, journalism, social media, telecommunications, and advertising. Some of the main opportunities in this domain include personalized recommendations for TV shows, movies, and music; improved user experience; more efficient content creation; improved communication and networking opportunities; increased brand exposure and sales; increased creativity and innovation; improved access to accurate information and fraud detection; and reduced costs. Using ChatGPT in this domain poses various threats, such as data privacy and security issues; ethical and legal concerns; fake news and media bias; cyberbullying and harassment; inauthenticity and lack of trust; the possibility of biased recommendations; and difficulty in monetizing created content.

\subsection{Arts and Culture}
The arts and culture domain encompasses a diverse range of applications, including music, art and design, fashion, writing and science fiction, gaming, virtual reality, performing arts, galleries, and social impact (social justice, environmental justice, activism, diversity, equity, inclusion, and social entrepreneurship). ChatGPT can provide numerous opportunities for improving processes and enhancing creativity in this domain. For example, it can facilitate improved music analysis, writing processes, artistic expression, fashion analysis, art curation, and gaming experiences. Additionally, ChatGPT can improve efficiency in composition and production, design processes, and content creation. While some studies have suggested that ChatGPT could potentially enhance certain cognitive functions, more research is needed to understand these effects better. In addition to privacy, ethical, and legal concerns, another specific threat in the arts and culture domain is the potential for biased recommendations to perpetuate stereotypes or reinforce existing inequalities. This could negatively affect the representation of diverse voices in the arts and culture domain. 

\subsection{Lifestyle and Leisure}
The lifestyle and leisure domain has applications in fitness and sports, entertainment, travel and hospitality, and food and beverage.
The main opportunities in this domain include improved fitness outcomes, increased motivation, enhanced performance analysis, efficient sports betting, improved fan engagement, better customer experience, increased sales, personalized recommendations for sports and travel, improved travel planning, improved product quality, and reduced costs.
The possible threats in this domain are privacy concerns and potential for biased recommendations.

\section{Experimental Study}
\begin{figure}[htb]
    \centering
    \includegraphics[scale=0.30]{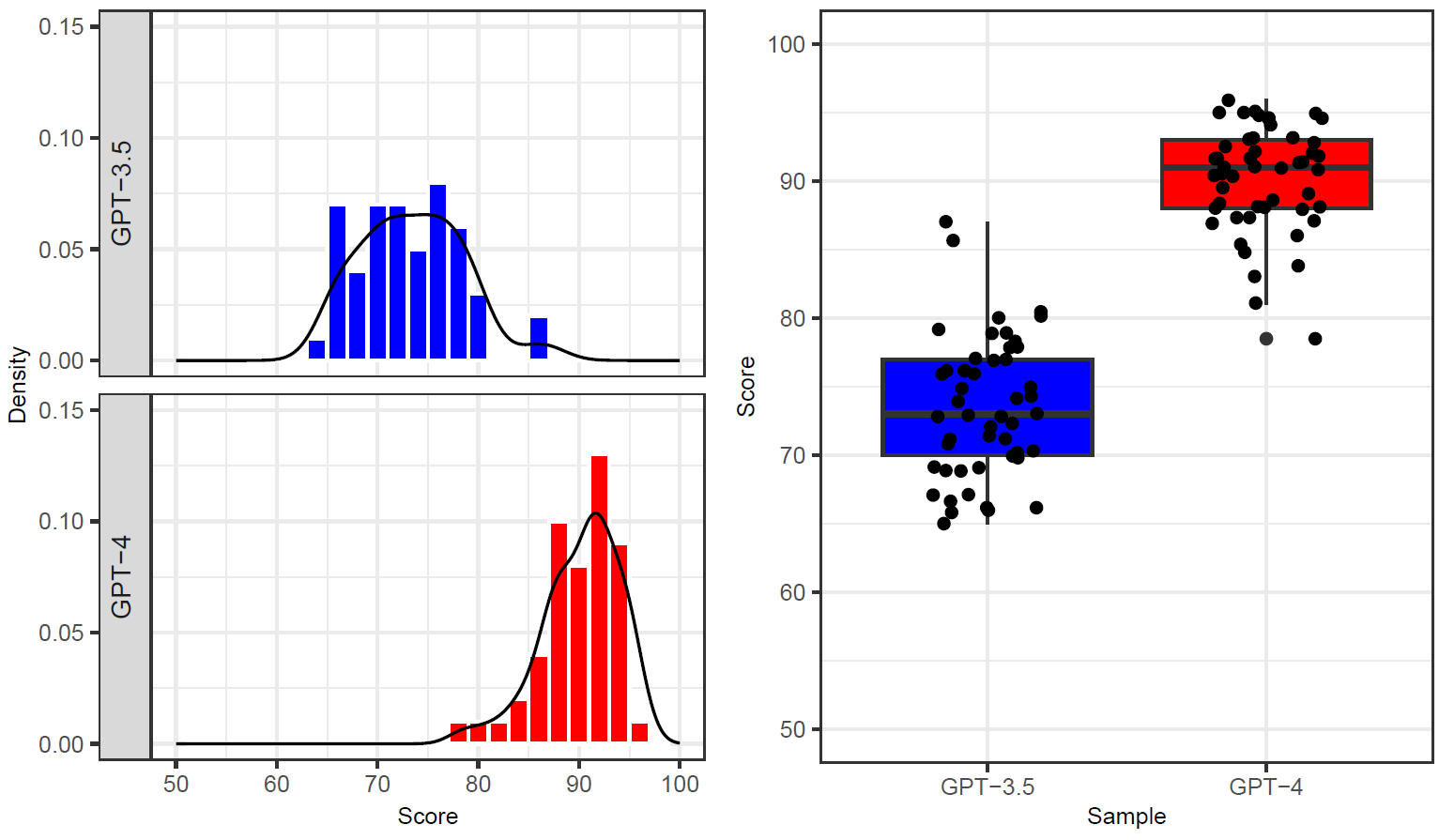}
    \caption{Histograms on the left and box plots on the right 
for GPT-3.5 and GPT-4 data}
    \label{fig1}
   \vspace{-0.60 cm}
\end{figure}

In this section, we focus on the domain of education discussed in section \ref{subedu} and perform an experimental study. Our case study is the midterm exam of SYS 3060: Stochastic Decision Models, a junior level course in the Department of Systems and Information Engineering at the University of Virginia, held in person in March 2023. As input in our experiment, we only told ChatGPT that these questions were related to the course topic. We gave the questions to both versions of GPT-3.5 and GPT-4 and repeated the experiment 50 times ($n_1=n_2=50$), and to prevent possible biases, we performed these experiments with different Internet Protocols (IPs), and the same person graded all the exams with the opportunity of receiving partial credits in 0.5 increments. Table \ref{Stat} summarizes the result of these experiments.
\begin{table}[htbp]
 \caption{Statistical properties of the experimental study of ChatGPT-3.5 and ChatGPT-4}
 \begin{center}
     \begin{tabular}{lcc}
    \hline
    ~ & \multicolumn{1}{l}{ChatGPT-3.5} & \multicolumn{1}{l}{ChatGPT-4}\\
    \hline
    Sample Size     & 50   & 50\\
    Mean  & 73.5 & 90.19 \\
    Standard Error of Mean & 0.72 & 0.54 \\
    Median & 73 & 91  \\
    Mode  & 76 & 88    \\
    Standard Deviation  & 5.05 & 3.84 \\
    Skewness & 0.37  & -0.86 \\
    Kurtosis & -0.09  & 0.77  \\
    Min   & 65 & 78.5    \\
    Max   & 87 & 96 \\
    Range & 22 & 17.5 \\
    \hline
    \end{tabular}
  \label{Stat}
 \end{center}
\end{table}
We conducted normality assumption tests with a significance level of $\alpha=0.05$ for GPT-3.5, and the Shapiro-Wilk and Anderson-Darling tests resulted in a $p$-value greater than $\alpha$, indicating that the data can be assumed to be approximately normal. However, for GPT-4, the data is not normal but negatively skewed, indicating that most scores are closer to the higher end and the maximum score. Since one dataset is assumed to be normally distributed and the other is not, instead of a t-test, we used the Mann-Whitney U test (also known as the Wilcoxon rank-sum test), which does not assume any particular distribution of the data. The result of the test showed a $p$-value very close to zero ($p$-value = $2.2 \times 10^{-16} < 0.05 = \alpha$), leading to the rejection of the null hypothesis, which asserts that the medians of the two samples are identical. Figure \ref{fig1} displays the histograms and box plots of the data for GPT-3.5 and GPT-4. We can observe that the median score of GPT-4 is significantly higher than that of GPT-3.5, indicating that GPT-4 outperforms the previous version.
It is worth noting that some of the questions on the exam were referring to subjects discussed in the class, and that is why none of the samples obtained by GPT-4 could reach the max score, which is 100\%.

\section{Conclusions}
This paper examines the evolution of ChatGPT and studies its applications, opportunities, and threats with a  focus on the business and industry and education domains. We conducted an experiment to evaluate the effectiveness of GPT-4 in an educational setting. While ChatGPT has revolutionized natural language processing and has the potential to save time and money by automating routine tasks, it is essential to be mindful of the potential threats and take steps to mitigate them, as it can produce misleading results and biases, raise ethical concerns, and be misused. That is why some countries, such as Italy, have banned the use of this AI \cite{banforbes,banbbc}, and responsible use of ChatGPT must be a top priority to ensure that everyone benefits from its use. 
While these models have shown impressive capabilities, we believe it is unlikely that they will be able to fully replace humans or individuals in all tasks and situations, and the jobs that are most likely to be affected are the ones that require routine and repetitive tasks \cite{ubcreport}. Machines may struggle to replicate human intuition, emotion, creativity, and intelligence, which are essential for many tasks; however, they can assist humans in developing new ideas and insights based on existing data and may be able to create entirely new concepts with human input and guidance.

\bibliography{refs}

\end{document}